\input{aipcheck}

\documentclass[
    ,final            
  ]
  {aipproc}

\layoutstyle{8x11double}

\usepackage{epsfig}
\usepackage{graphicx,bbm}

\newcommand{\be}{\begin{equation}}
\newcommand{\ee}{\end{equation}}
\newcommand{\bea}{\begin{eqnarray}}
\newcommand{\eea}{\end{eqnarray}}

\def\<{\left\langle}
\def\>{\right\rangle}
\def\be{\begin{equation}}
\def\ee{\end{equation}}
\def\bea{\begin{eqnarray}}
\def\eea{\end{eqnarray}}
\newcommand{\lwig}{\mbox{\,\raisebox{.3ex}
    {$<$}$\!\!\!\!\!$\raisebox{-.9ex}{$\sim$}\,}}

\begin{document}

\title{The Flavour of Inflation}


\author{I.~Zavala}{
  address={
IPPP and CPT, Durham University, DH1 3LE, Durham, UK}
}

\begin{abstract}
  A new class of particle physics models of inflation based on the      
phase transition associated with the spontaneous breaking of family symmetry  is proposed. The Higgs fields responsible for the breaking of  family symmetry, the flavons, are natural inflaton candidates  or waterfall fields in hybrid inflation. This opens up a rich vein of possible inflation models, all linked to the physics of flavour, with several interesting cosmological implications.   
\end{abstract}

\maketitle



The inflationary paradigm was proposed more that 20 years ago 
\cite{Guth:1980zm}. Yet, it has been only recently that its predictions have been firmly demonstrated  to be consistent with observations \cite{wmap5}.
In spite of this success, to date, the precise origin  of the inflation field(s) and its relation to particle physics remains obscure, despite much effort done in this direction \cite{reviews}. 

On the other hand, one of the greatest mysteries facing the Standard Model (SM) of particle physics 
is the  
origin of flavour structure of quarks and leptons, its masses and mixing angles.
A common approach to understand this flavour structure, is to assume that the SM is extended by some  horizontal family symmetry  $G_F$ \cite{gflsym}. This symmetry  may be continuous
or discrete, and gauged or global. It must be broken completely,
apart from possibly remaining discrete symmetries,
at some high energy scale in order to be phenomenologically
consistent. Such  symmetry breaking requires the introduction
of new Higgs fields called {\em flavons}, $\phi$, whose vacuum 
expectation values (vevs) break the family symmetry $\<\phi \>\neq 0$.
In addition, most of the family symmetry models rely on supersymmetry (SUSY) where besides the flavon superfields, additional ``driving''  superfields (usually singlets under $G_F$), associated with the vacuum alignment of the flavons, are present. 
 
In this talk, we suggest that the phase transition 
associated with the spontaneous breaking of family symmetry
is also responsible for cosmological inflation, a possibility 
we refer to as {\em flavon inflation} \cite{akmvsz}.
In other words, we propose to  use flavour physics as a guideline for models of inflation and inflation to provide us with constrains on the flavour physics parameters.

Working in a supergravity framework, there are at least two possible flavon inflation scenarios that can arise: either  the flavon fields  play the role of the inflaton, or the driving fields play the role of the inflatons, whereas the flavons fields  end inflation as waterfall fields in a (SUSY)  hybrid inflation model. In what follows we present an example of each of these models\footnote{ 
A third scenario, which we do not consider here, is where both the flavon fields and the driving fields can play the role of inflatons in a multistage, multifield type of model \cite{progress}.}.

Consider the situation where the inflatons are representations of a family symmetry group.  For
example  a non-Abelian discrete family symmetry $A_4\in SU(3)$,
as in several recent models of lepton (neutrino) mass structures \cite{a4}. In this case, combinations of the form $\phi_1 \phi_2 \phi_3$ can arise since for fields $\phi =
(\phi_1,\phi_2,\phi_3)$, $\psi = (\psi_1,\psi_2,\psi_3)$ and $\chi =
(\chi_1,\chi_2,\chi_3)$ each in the fundamental triplet $\underline{3}$
representation of $A_4$,  $\{\phi_1 \psi_2 \chi_3$ + permutations$\}$ forms an invariant. 

\begin{figure}
  \includegraphics[height=.24\textheight]{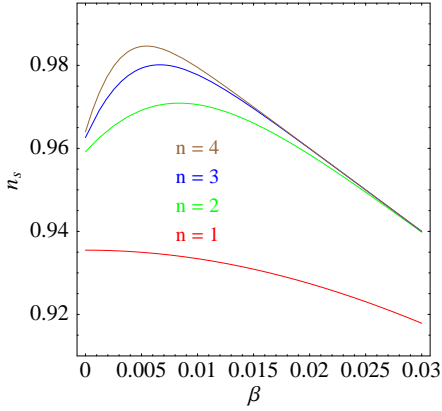}
  \caption{Predictions for the spectral index $n_s$ as a function of $\beta$ for $N=60$.}
\label{fig:nsresults}
\end{figure}

For concreteness, let us look at the following toy model where the superpotential takes the form (this is similar to \cite{Senoguz:2004ky})
\begin{eqnarray}\label{Eq:InvHybrid}
W = S \left[ \frac{(\phi_1 \phi_2 \phi_3)^n}{M_*^{3n-2}} - \mu^2  \right] \,;
	\qquad  	\qquad      n \ge 1
\end{eqnarray} 
and the non-minimal K\"ahler potential 
\begin{eqnarray}
 K = |S|^2 + |\phi|^2 +  \frac{ \kappa_2\,|S|^2 |\phi|^2}{M_P^2} + 
\frac{ \kappa_1\,|\phi|^4}{4 M_{Pl}^2} +\frac{ \kappa_3\,|S|^4}{4 M_{Pl}^2} + ... 
\label{eq:inf_flav_A4}
\end{eqnarray}
where $M_{Pl} = (8\pi G_N)^{-1/2}\sim 10^{18}$GeV is the reduced Planck mass. The supergravity F-term scalar potential\footnote{Since
 $A_4$ is a discrete symmetry, there are no D-term contributions                               
to the potential. }  can be calculated from the standard relation
\be\label{sugrapot} 
V= e^{K/M_{Pl}^2} \, \left[ K^{i\bar j}D_{i}W D_{\bar j}\overline{W} - 3\frac{|W|^2}{M_{Pl}^2}
\right] \,.
\ee
with  $D_iW = \partial_iW + M^{-2}_{Pl} W\partial_i K$. The global minimum is found from  $D_{i} W=0$ and $W=0$, to be located at $\phi_i \sim M$, where $M = M_*(\mu/M_*)^{2/3n}$ is the symemtry breaking scale, and  $S=0$. For cosmological  porpouses, we work in the limit $S\ll M_{Pl}$ and $\phi_i \ll M \ll M_{Pl}$. 
We look at the inflationary trajectory where  $S$ picks a large mass and sits at its minimum at zero (this implies $\kappa_3 < -1/3$), whereas the flavon fields roll slowly from zero, where the family symmetry is restored, towards their true minimum. 
Defining  the  canonically normalised fields as $|\phi_i| \equiv \varphi_i /\sqrt{2}$ and
concentrating  in the generic trajectory $\varphi_1 \sim \varphi_2\sim \varphi_3 \sim \varphi$, the scalar potential during inflation becomes 
 \be 
 V \simeq \mu^4 \left[ 1 -
\frac{\beta}{2} \frac{\varphi^2}{ M_{Pl}^2}  - \gamma
\frac{\varphi^{3n}}{M^{3n}} + \cdots\right] \,,
\ee 
where $\beta = (\kappa_2 -1)$ and  $\gamma = 2/(6)^{3n/2}$.
Using the standard slow roll inflationary parameters, $(\epsilon, \eta)$, the spectral index 
$n_s -1 = 2\eta -6\varepsilon$,  can be expressed in terms of the number of e-folds $N$ as (in the present model $\epsilon \ll \eta$)
 $$ n_s \approx 1 - 2 \beta \left[ 1+
\frac{(3n-1)(1-\beta)} {[(3n-2)\beta + 1] e^{\beta(3n-2)N} +\beta -1}
\right]\,,  $$ %
for $\beta \neq 0 $ and 
$$n_s \approx 1 -  \frac{6n-2}{(3n-2)N + (3n-1)} \,,
$$
for $\beta =0$.
The results are illustrated in figure \ref{fig:nsresults}. The predictions for $n_s$ are close to the WMAP5 year data \cite{wmap5} $n_s = 0.960 \pm 0.014$, for $n\ge 2$ and $\beta \lwig 0.03$ (taking $N = 60$).

\begin{figure}
  \includegraphics[height=.5\textheight]{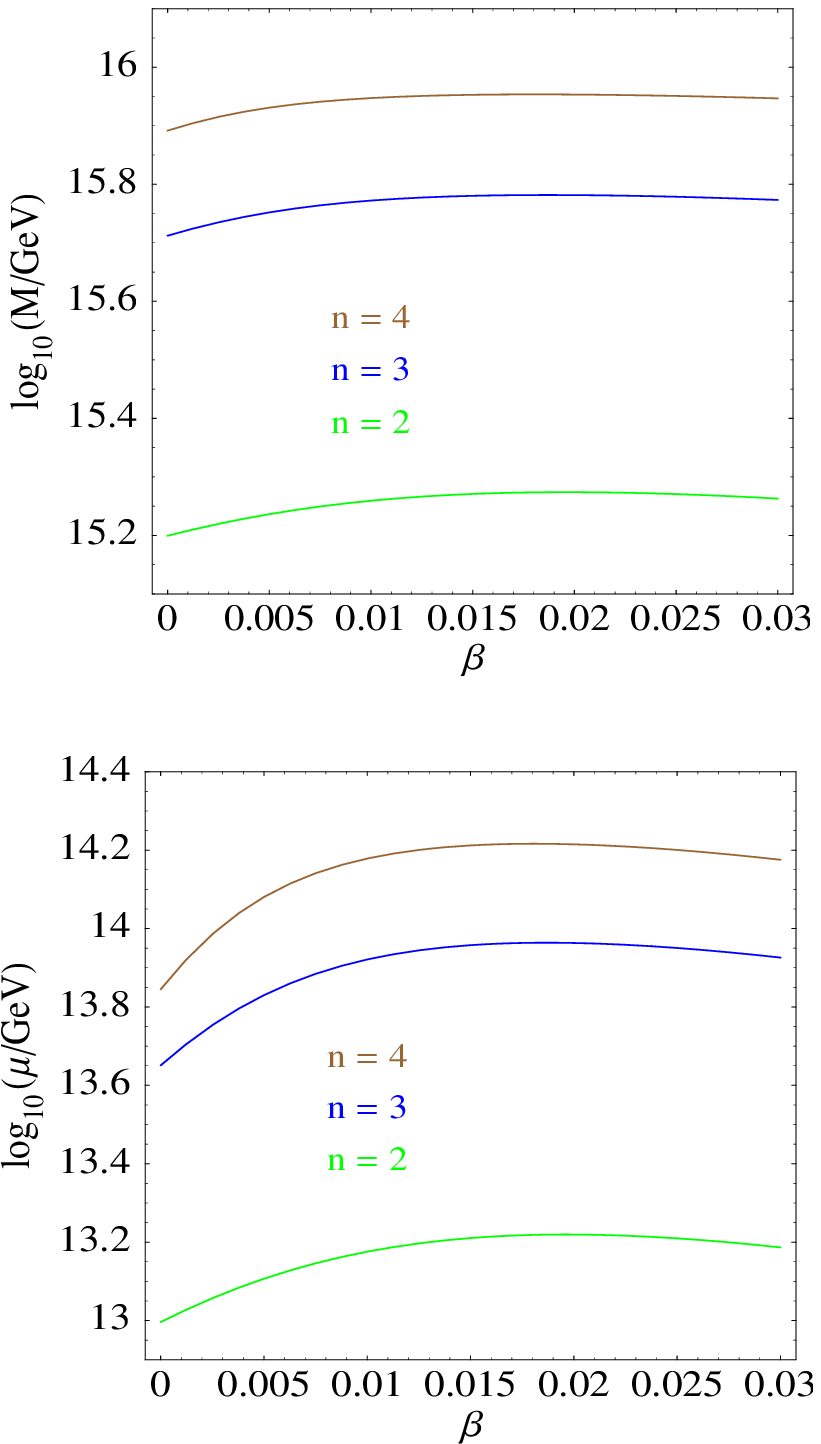}
  \caption{Predictions for the family breaking scale $M$ and the  inflation scale $\mu$ as a function of $\beta$ for $N=60$ and $M_* = M_\mathrm{GUT}$.}
\label{fig:MandMuresults}
\end{figure}

We can relate the scale $M_*$ to the family symmetry breaking scale $M$ and the inflation scale $\mu$, via the amplitude of the density perturbation when it re-enters the horizon 
$$\delta_H = \frac{2}{5} {\mathcal P}_{\mathcal R}^{1/2} = 
\frac{1}{5\pi\sqrt{3}} \frac{V^{3/2}}{M_{Pl}^3|V'|}  = 1.91 \times 10^{-5}\,.$$
With $M_*$ around the GUT scale ($M_{GUT} = 2\times 10^{16}$GeV), we obtain 
  $M \approx (10^{15},10^{16})  \;\mbox{GeV}$ for $n=2,3,4$ 
and $\approx (10^{13},10^{14})  \;\mbox{GeV}$ for $n=2,3,4$, 
for all considered values of $\beta$.
The numerical results are shown in figure \ref{fig:MandMuresults}.
Note that the choice $M_* = M_\mathrm{GUT}$ is only an example. In principle $M_*$ can be much lower and thus also $M$ would be much lower (for example an intermediate scale $M_*  \approx (10^{11},10^{13}) \;\mbox{GeV}$ can give $M\approx (10^{11},10^{13})\;\mbox{GeV} $, $\mu \approx (10^{10},10^{12})\;\mbox{GeV}$).

\smallskip

The second scenario we consider has  the driving fields (singlets under the family  symmetry) acting as inflatons, whereas the flavon fields  are the waterfall fields  in a susy hybrid inflation. 
For this example we consider an explicit $SU(3)$ family symmetry model, already studied in the literature \cite{deMedeirosVarzielas:2005ax}, where the symmetry breaking occurs in steps. For concreteness, we assume that it is in the last step that inflation takes place\footnote{A more general situation can arise if at each stage of family symmetry breaking, a stage of inflation also takes place, giving rise to  a multistage flavon inflationary scenario \cite{progress}. }. 

The relevant part of the superpotential which governs the final step of family symmetry breaking is given by 
\begin{eqnarray}
 W &=& \kappa S (\bar\phi_{123} \phi_{123} - M^2) + \kappa' Y_{123} \bar\phi_{23} \phi_{123} \nonumber \\
&& \hskip2cm 
+ \kappa'' Z_{123} \bar\phi_{123} \Sigma \phi_{123} + ... \;.
\label{mvr_w23}
\end{eqnarray}
where the flavon fields $\<\phi_{23} \> \propto (0,1,1)^T$ and $\<\Sigma \>= \mbox{diag}(a,a,-2a)$ have already acquired a vev and are 
already sitting at their minima \cite{deMedeirosVarzielas:2005ax}.
 $S$ is the driving superfield for the flavon $\phi_{123}$, i.e.\ the contribution to the scalar potential from $|F_S|^2$ governs the vev $\< \phi_{123} \>$.   
In addition, we take the non-minimal K\"ahler potential: 
\bea\label{kahler1}
K&=&
|S|^2 + |\phi_{123}|^2 + |\bar\phi_{123}|^2 + |Y_{123}|^2 + |\bar\phi_{23}|^2 + |\phi_{23}|^2 \nonumber \\
&&
+ |Z_{123}|^2  
+ \kappa_S\frac{|S|^4}{4 M_{Pl}^2}
+ \kappa_{SZ}\frac{|S|^2 |Z_{123}|^2}{4 M_{Pl}^2}
+ ...
\;.
\eea
The inflationary trajectory occurs when the fields with larger vevs 
$Y_{123},\,\phi_{123},\,\bar\phi_{123}$ do not evolve, whereas those with smaller vevs,  $Z_{123}$ and $S$  move towards their minimum at zero.  
Focusing on the D-flat directions  
and setting $Y_{123}=\phi_{123}=\bar\phi_{123}=0$ since they  obtain large masses from the superpotential, the tree level scalar potential during inflation takes the  form\footnote{The dots include also quartic terms  such that the fields  evolve from large to small values. } 
\bea
V =  \kappa^2 M^4\,   \left[ 1 -  \gamma \frac{\xi^2}{2M_{Pl}^2}   
- 2\kappa_{S} \frac{\sigma^2}{2 M_{Pl}^2} + ...\right],
      \label{sugrapoteff1}
\eea
where $|S| = \sigma/\sqrt{2}$, $|Z_{123}| = \xi/\sqrt{2}$ and $\gamma = \kappa_{SZ}-1$. 
If the two coefficients in front of the mass terms for $\sigma$ and/or $\xi$ are sufficiently small these can drive inflation. 
In the single field case \cite{BasteroGil:2006cm}, where $\sigma$ acts as the inflaton (choosing for instance $\gamma < -1/3$ such that the mass of $\xi$ exceeds the Hubble scale) and taking into account loop corrections to the potential \cite{lr}, one finds that  for $\kappa_{S} \approx (0.005 - 0.01)$ and $\kappa \approx (0.001 - 0.05)$, a spectral index consistent with WMAP5 year data \cite{wmap5}, $n_s = 0.96 \pm 0.014$, is obtained \cite{BasteroGil:2006cm}. 
Finally, the scale $M$ of family symmetry breaking along the $\< \phi_{123} \>$-direction is determined 
from the temperature fluctuations $\delta T/T$ of the CMB to be
$M \approx 10^{15} \;\mbox{GeV},$
about an order of magnitude below the GUT scale. 

In summary, we proposed that there exists  a non-trivial connection between the physics of flavour and that of inflation. The flavon fields responsible to break family symmetries, can act as inflatons or as waterfall fields in a hybrid inflation scenario, with the driving fields of flavour symmetries acting as inflatons. 
Because family symmetries have to be completely broken, flavon inflation avoids formation of unwanted relics like those associated to the breaking of GUT symmetries.
Moreover, the symmetry breaking scale can be linked to that of inflation, thus making inflation a source of phenomenological information for the flavour sector of particle physics. 
We have shown briefly how this connection can arise, but it is clear that further relations between the parameters of inflation and those of the family symmetry structure can exist. 
For example, a detailed  study of  reheating, baryogenesis and (non-thermal) leptogenesis can be expected to reveal a more precise connection between these two sectors \cite{progress}. 

As a last remark, we point out that in the present proposal we did not intended to provide a solution to the  well known $\eta$ problem  
in theories of supegravity inflation. We have taken a more phenomenological approach, being more interested in the physics connecting family symmetries and inflation. 
It is possible to imagine, however,  flavon inflation scenarios which avoid this problem, for example D-term flavon inflation and others \cite{progress}.

\begin{theacknowledgments}
  This talk is based on work done in collaboration with S.~Antusch, S.~King, M.~Malinsky and L.~Velasco-Sevilla. The author also thanks G.~Ross, Q.~Shafi and G.~Tasinato for discussions. This work was supported by an STFC Postdoctoral Fellowship. 
\end{theacknowledgments}


\begin{thebibliography}{99}
 
\bibitem{Guth:1980zm}
  A.~H.~Guth,
  Phys.\ Rev.\  D {\bf 23} (1981) 347.

\bibitem{wmap5}
  E.~Komatsu {\it et al.}  [WMAP Collaboration],
  arXiv:0803.0547 [astro-ph].

\bibitem{reviews}
  See for example:  D.~H.~Lyth and A.~Riotto,
  Phys.\ Rept.\  {\bf 314} (1999) 1
  [arXiv:hep-ph/9807278].

\bibitem{gflsym} 
  C.~D.~Froggatt and H.~B.~Nielsen,
  Nucl.\ Phys.\  B {\bf 147} (1979) 277;
  ~J.~L.~Chkareuli, C.~D.~Froggatt and H.~B.~Nielsen,
  Nucl.\ Phys.\  B {\bf 626}, 307 (2002)
  [arXiv:hep-ph/0109156].
 
 
\bibitem{akmvsz}  
  S.~Antusch, S.~F.~King, M.~Malinsky, L.~Velasco-Sevilla and I.~Zavala,
  arXiv:0805.0325 [hep-ph].

   
 \bibitem{progress}
Work in progress.


\bibitem{a4}
  I.~de Medeiros Varzielas, S.~F.~King and G.~G.~Ross,
  Phys.\ Lett.\  B {\bf 644} (2007) 153
  [arXiv:hep-ph/0512313];
  Phys.\ Lett.\  B {\bf 648} (2007) 201
  [arXiv:hep-ph/0607045];
  E.~Ma,
  Mod.\ Phys.\ Lett.\  A {\bf 21} (2006) 2931
  [arXiv:hep-ph/0607190];


%
  
\bibitem{Senoguz:2004ky}
  V.~N.~Senoguz and Q.~Shafi,
  Phys.\ Lett.\  B {\bf 596} (2004) 8
  [arXiv:hep-ph/0403294].





\bibitem{deMedeirosVarzielas:2005ax}
  I.~de Medeiros Varzielas and G.~G.~Ross,
  Nucl.\ Phys.\  B {\bf 733} (2006) 31
  [arXiv:hep-ph/0507176].


   

\bibitem{lr}
  A.~D.~Linde and A.~Riotto,
  Phys.\ Rev.\  D {\bf 56} (1997) 1841
  [arXiv:hep-ph/9703209].

\bibitem{BasteroGil:2006cm}
See for instance:  M.~Bastero-Gil, S.~F.~King and Q.~Shafi,
  Phys.\ Lett.\  B {\bf 651} (2007) 345
  [arXiv:hep-ph/0604198].




 

\end{thebibliography}
\end{document}